\documentclass[conference]{IEEEtran}
\ifCLASSINFOpdf
\else
\fi
\usepackage{url,doi}

\usepackage{epsfig}
\usepackage{epstopdf}
  \DeclareGraphicsExtensions{.pdf,.jpeg,.png,.eps}

\hypersetup{%
linkcolor=blue,anchorcolor=blue,%
citecolor=blue,filecolor=blue,%
menucolor=blue,%
urlcolor=blue,%
colorlinks=true}

\begin{document}
%
\title{Preconditioned Spectral Clustering for\\Stochastic Block Partition\\Streaming Graph Challenge}

\author{\IEEEauthorblockN{David Zhuzhunashvili}
\IEEEauthorblockA{University of Colorado\\
Boulder, Colorado\\
Email: David.Zhuzhunashvili@colorado.edu}
\and
\IEEEauthorblockN{Andrew Knyazev}
\IEEEauthorblockA{Mitsubishi Electric Research Laboratories (MERL)\\
201 Broadway, 8th Floor, Cambridge, MA 02139-1955\\
Email: knyazev@merl.com, WWW: \url{http://www.merl.com/people/knyazev}}
}


%

\IEEEspecialpapernotice{(Preliminary version at arXiv.)}

\maketitle

\begin{abstract}
Locally Optimal Block Preconditioned Conjugate Gradient (LOBPCG) is demonstrated to efficiently solve eigenvalue problems for graph Laplacians that appear in spectral clustering. For static graph partitioning, 10--20 iterations of LOBPCG without preconditioning result in \~{}10x error reduction, enough to achieve 100\% correctness for all Challenge datasets with known truth partitions, e.g., for graphs with  5K/.1M (50K/1M) Vertices/Edges in 2 (7) seconds, compared to over 5,000 (30,000) seconds needed by the baseline Python code. Our Python code 100\% correctly determines 98 (160) clusters from the Challenge static graphs with 0.5M (2M) vertices in 270 (1,700) seconds using 10GB (50GB) of memory. Our single-precision MATLAB code calculates the same clusters at half time and memory. For streaming graph partitioning, LOBPCG is initiated with approximate eigenvectors of the graph Laplacian already computed for the previous graph, in many cases reducing 2-3 times the number of required LOBPCG iterations, compared to the static case.  Our spectral clustering is generic, i.e. assuming nothing specific of the block model or streaming, used to generate the graphs for the Challenge, in contrast to the base code. Nevertheless, in 10-stage streaming comparison with the base code for the 5K graph, the quality of our clusters is similar or better starting at stage 4 (7) for emerging edging (snowballing) streaming, while the computations are over 100--1000 faster.
\end{abstract}


%
\IEEEpeerreviewmaketitle

\section{Introduction}\label{s:bm}

Spectral clustering is a classical method for grouping together relevant data points, while at the same time separating irrelevant data points. Spectral clustering is commonly formulated as a partitioning of a graph, with vertices of the graph being mapped with data points. The clustering is called ``spectral'' because its algorithms are based on spectral graph theory, i.e. spectral properties of matrices associated with the graph, such as graph adjacency and Laplacian matrices, and thus can be related to dimensionality reduction via the principal component analysis. Compared to other practical clustering techniques, spectral clustering is arguably well supported mathematically, mostly avoiding heuristic simplifications, typically needed in combinatorial optimization formulations of clustering to make computations tractable. 
Spectral clustering has been successful in a wide variety of applications, ranging from traditional resource allocation, image segmentation,  and information retrieval, to more recent bio-informatics, providing meaningful results at reasonable costs. 

The graph partitioning problem can be formulated in terms of spectral graph theory, e.g., using a spectral decomposition of a graph Laplacian matrix, obtained from a graph adjacency matrix with non-negative entries that represent positive graph edge weights describing similarities of graph vertices. Most commonly, a multi-way graph partitioning is obtained from approximated ``low frequency eigenmodes,'' i.e. eigenvectors corresponding to the smallest eigenvalues, of the graph Laplacian matrix. Alternatively and, in some cases, e.g.,\ normalized cuts, equivalently, one can operate with a properly scaled graph adjacency matrix, turning it into a row-stochastic matrix that describes probabilities of a random walk on the graph, where the goal is to approximate the dominant eigenpairs. 

The Fiedler vector, or a group of eigenvectors of the graph Laplacian corresponding to the left-most eigenvalues, are computed iteratively by solving the corresponding symmetric eigenvalue problem. Efficiency of iterative solvers is crucial for practical computations for large or real time streaming data.
The need to deal with big data and the resulting humongous matrix eigenvalue problems in data sciences is not historically the first one. Computational mechanics and especially material sciences have long been sources of large scale eigenvalue problems, with the increasing needs outgrowing all advances in computing resources. For example, a decade ago two Gordon Bell Prize finalists at ACM/IEEE Conferences on Supercomputing in 2005 \cite{1559996} and 2006 \cite{Yamada:2006:HCE:1188455.1188504} implemented on Japan's Earth Simulator---the number one supercomputer in the world at that time,---and successfully tested, for multi-billion size matrices, the Lanczos~\cite{Lanczos50aniterative} and Locally Optimal Block Preconditioned Conjugate Gradient (LOBPCG)~\cite{K01} methods. 

The remainder of the paper is organized as follows. In \S\ref{s:p}, we briefly review, cf.~\cite{Luxburg2007,chapter_sc}, traditional graph-based spectral clustering and cuts partitioning, discuss numerical issues of related large scale computations for Big Data spectral clustering, and describe our approach. Partition Challenge Datasets with Known Truth Partitions \cite{GC17} suitable for our techniques are reviewed in \S\ref{s:e}. Our numerical results appear~in~\S\ref{s:r}. 
We~conclude in \S\ref{s:c} that spectral clustering via LOBPCG is an efficient scalable technology for high-quality fast graph partitioning in the present era of Big Data.

\section{Approach}\label{s:p}%
\subsection{Introduction to spectral clustering in a nutshell}\label{ss:sc}%

Let entries of the matrix $W$ be called \emph{weights} and the matrix $D$ be diagonal, made of row-sums of the 
matrix $W$. The matrix $W$ may be viewed as a matrix of scores, which represent the similarity between of data points. Similarities are commonly determined from their counterparts, distances. The distance matrix is a matrix containing the distances, taken pairwise, of a set of data points. The general connection is that the similarity is small if the distance is large, and vice versa. 

Commonly, the data clustering problem is formulated as a graph partition problem. The graph partition problem is defined on data represented in the form of a graph $G = (V, E)$, with $n=|V|$ vertices $V$ and $m=|E|$ edges $E$ such that it is possible to partition $G$ into smaller components with specific properties. For instance, a $k$-way partition splits the vertex set into $k$ non-overlapping subsets. The similarity matrix $W$ is provided as an input and consists of a quantitative assessment of the relative similarity of each pair of points in the dataset. In the framework of graph spectral partitioning, entries of the $n$-by-$n$ matrix $W$ are weights of the corresponding  edges $E$, and the matrix $W$ is called the \emph{weighted graph adjacency} matrix.

Traditional mathematical definitions of graph partitioning are combinatorial and naturally fall under the category of NP-hard problems, solved using heuristics in practice. 
Data clustering via graph spectral partitioning , initiated in \cite{fiedler1973algebraic,fiedler1975property}, 
is a state-of-the-art tool, which is known to produce high quality clusters at reasonable costs. 

In directed graphs, the weighted graph adjacency matrix $W$ is not symmetric. For the purpose of spectral partitioning, which is based on eigenvalue decompositions of matrices associated with the graph, it is very convenient to deal with symmetric matrices, so the non-symmetric $W$ is often symmetrized. The symmetrization we use is substituting $W+W'$ for $W$, where 
$W'$ denoted the transpose of the matrix $W$ and the sum ``$+$'' may be logical if the graph has no weights, e.g., all non-zero entries in $W$ are simply ones. In general, we~assume that all the entries (weights) in $W$ are non-negative real numbers and that the sum is algebraic.  

Let us introduce the graph Laplacian matrix $L=D-W$.
The column-vector of ones is always an eigenvector of $L$ corresponding to the zero eigenvalue. 
The symmetric matrix $L$ has all non-negative eigenvalues,
provided that all entries of the matrix $W$ are nonnegative; cf. \cite{2017arXiv170101394K}.
The actual number of  major partitions is automatically determined in spectral clustering using a gap in the smallest eigenvalues of the graph Laplacian. The gap is calculated by comparing the increments in the eigenvalues to find the first significantly large increment.  The number of eigenvalues located below the gap determines the number of major multi-way graph partitions/clusters. An~absence of the gap indicates that no reliable multi-way graph partitioning is available. Multi-way block partition is computed from the matrix of approximate eigenvectors via QR factorization with column pivoting, shown in 
\cite{2016arXiv160908251D} to supplement or supersede the traditional $k$-means++.

\subsection{Numerical challenges of spectral clustering}\label{ss:nc}%

Our spectral clustering performs block partitioning of graphs using the Locally Optimal Block Preconditioned Conjugate Gradient (LOBPCG) method to iteratively approximate leading eigenvectors of the symmetric graph Laplacian for multi-way graph partitioning. The number of the eigenvectors should not subceed the number of anticipated partitions. 

If further lower-level partitioning is desired, the multi-way clustering procedure described above can be repeated recursively to independently partition each of the already determined clusters, resulting in a hierarchical tree of block partitions.  Our code, written for the IEEE HPEC Graph Challenge 2017 \cite{GC17}, does not have this feature implemented, performing only a single multi-way graph partitioning.

Both Lanczos and LOBPCG methods do not require storing a matrix of the eigenvalue problem in memory, but rather only need the results of multiplying the matrix by a given vector. Such a matrix-free characteristic of the methods makes them particularly useful for eigenvalue analysis problems of very large sizes, and results in good parallel scalability for large-size matrices on multi-threaded computational platforms with many parallel processors and cores. 

Compared to Lanczos, LOBPCG is a block method, where several eigenvectors are computed simultaneously as in the classical subspace power method. Blocking is beneficial if the eigenvectors to be computed correspond to clustered eigenvalues, which is a typical scenario in multi-way spectral partitioning, where often a cluster of the smallest eigenvalues is separated by a gap from the rest of the spectrum. Blocking also allows taking advantage of high-level BLAS3-like libraries for matrix-matrix  
operations, which are typically included in CPU-optimized computational kernels.

LOBPCG may be optionally accelerated by preconditioning, if available, e.g.,\ see \cite{K01,k2003,BLOPEX}, to speed up convergence, e.g., using Algebraic Multigrid Method (AMG) applied to a regularized graph Laplacian. The regularization is needed, since AMG attempts to mimic an action of a matrix inverse, while the original graph Laplacian is technically not invertible, always having the trivial constant eigenvector corresponding to the zero eigenvalue. 
AMG preconditioning increases the computational costs per iteration, but may reduce the number of required LOBPCG iterations.  AMG preconditioning is known to perform extremely well for some classes of graphs, although may be inefficient for arbitrary graphs. Since  preconditioning is optional in LOBPCG, it can be dynamically turned off or on, as well as AMG parameters may be tuned, depending on past performance in the case of streaming graph. 

The open-source Python function of LOBPCG from SciPy is used for our numerical tests; cf.,\ \cite{DBLP:journals/corr/McQueenMVZ16}. AMG and dense matrix factorizations, e.g., QR, are also performed by well-known open-source Python functions. Similar open-source LOBPCG and AMG functions are available in C with MPI and OpenMP parallel functionality, e.g., in hypre and PETSc/SLEPc \cite{BLOPEX}, Trilinos/Anasazi \cite{boman2014installing} libraries, as well as GPU implementations in MAGMA \cite{atd16} and NVIDIA nvGRAPH \cite{NaumovnvGRAPH2016}. 

\section{Experiments}\label{s:e}
IEEE HPEC Streaming Graph Challenge Stochastic Block Partition seeks to identify optimal blocks (or clusters) in a graph. In static processing, given a large graph $G$ the goal is to partition $G$.  In stateful streaming, which is the focus of the streaming graph challenge, given an additional smaller graph $g$ the goal is to partition $G + g$; see \cite{GC17}.  

Taking advantage of the second author expertise in numerical solution of matrix eigenvalue problems, we have selected the graph partition challenge, for which spectral clustering is known in practice to produce high quality clusters at reasonable computational costs. The costs of spectral clustering are primarily determined by a choice of an eigenvalue solver.  
All eigenvalue solvers are by nature iterative, so one needs to minimize the costs per iteration as well the total number of iterations needed to achieve the required accuracy of clustering. The LOBPCG eigenvalue solver, that we have selected to use, only requires the result of single multiplication of the graph adjacency matrix by a vector (or a group of vectors that is called a multi-vector) per iteration, just as the classical power method.    

For static graphs, the initial approximations for the eigenvectors of the graph Laplacian in LOBPCG are chosen randomly---every component of every eigenvector is Gaussian (alternatively, uniformly distributed on [-1,1]). Such randomization helps LOBPCG to avoid local stationary points in the iterative process of convergence to the leading eigenvectors. Streaming graphs are treated via warm-starts of LOBPCG, where the approximate eigenvectors already computed for the previous graph in the stream serve as high-quality initial approximations in LOBPCG for the graph Laplacian of the current graph. Such warm-starts of LOBPCG may significantly reduce the number of iterations of LOBPCG, compared to the randomized initialization for the static case, if the streaming graph is sampled often enough. To enable possible appearance of additional partitions of the streaming graph one should choose the starting number  of the eigenvectors to be computed large enough and keep it, or determine the number of computed eigenvectors dynamically, always slightly exceeding (for the gap calculation) the expected number of partitions. 

Streaming graphs with dynamically changing numbers of vertices, like in the snowball example, can also be efficiently processed by LOBPCG with warm-starts, if a mapping is known between the sets of old and new vertices. In the snowball example, the old vertices are always numbered first and in the same order in a subset of the new vertices, thus for warm-starts we put the previously computed eigenvectors in the top block of the initial approximate eigenvectors for the current graph. Since the set of vertices in the snowball example is growing, warm-starts require filling the missing entries in the bottom block of the initial approximate eigenvectors for the current graph. One can fill in the missing entries by zeros, randomly, or use interpolation of the already available values in the old vertices, e.g.,\ by propagating the values using the graph connectivity. In our tests, we use the zero fill-in.

We have limited our testing to Partition Challenge Datasets with Known Truth Partitions. In terms of the number of vertices of the graphs, our MATLAB, Python, and C codes partition the Challenge static 5K graph 100\% correctly in a second, already making reliable timing difficult, so 5K is the smallest size we have tested. Our Python code runs out of 128~GB memory in the case of 5M vertices with 221 partitions, so our largest tested Challenge graph is with 2M vertices. 

There are public eigensolvers in C with MPI, allowing well-scalable parallel distributed-memory solution of eigenvalue problems, e.g., LOBPCG in hypre and PETSc/SLEPc \cite{BLOPEX}, Trilinos/Anasazi \cite{boman2014installing}, and MAGMA \cite{atd16}. We have performed preliminary tests with LOBPCG in hypre and PETSc/SLEPc/BLOPEX on a single node, but observe no dramatic increase of performance, compared to our Python and MATLAB codes---so we do not report the corresponding results in the present work. Our~explanation is that, to our knowledge, the present versions of LOBPCG in hypre and SLEPc/BLOPEX implement the distributed column-matrix multi-vector $X$ only rudimentarily, as a collection of distributed single vectors, which does not allow utilizing highly efficient matrix-matrix BLAS-3 type libraries. In contrast, PETSc/SLEPc, Trilinos/Anasazi, and MAGMA versions of LOBPCG pay special attention to efficient implementation of distributed multi-vectors and thus are expected to perform well both in shared and distributed memory environments.

Our Python and MATLAB codes do not use MPI, so all our tests are on a single node. However, the BLAS and LAPACK-type functions in Python and MATLAB are multi-threaded and the most expensive operations in LOBPCG are matrix-matrix BLAS-3 type, so multiple cores are being implicitly utilized. 

We test matrix-free versions of our codes, where the original non-symmetric adjacency matrix $A$ is read from the given file and stored in binary sparse matrix format, and where the required by LOBPCG multiplication of the graph Laplacian $L$ by a column-matrix (block of vectors) $X$ is performed by a function call, e.g., in MATLAB notation,   
$LXfun = @(X)diagL.*X-A*X-(X'*A)'$, wherein the vector $diagL$ contains pre-computed diagonal entries of $L$. AMG commonly needs the matrix $L$ of the graph Laplacian being explicitly available, however. We find that our  matrix-free implementations in Python and MATLAB lead to no decrease in computing time, although saves memory for matrix storage. 

The Python AMG has not led to acceleration in our tests, and the corresponding results are not shown. We have also tested LOBPCG in hypre and PETSc/SLEPc with hypre algebraic multigrid BoomerAMG, but observed no advantages, so we report no results here.  Our explanation is that LOBPCG has been able to determine 100\% correct partition typically after a relatively small, 10--20, number of iterations, even without AMG. The extra costs of constructing and applying AMG preconditioning that we have tested are significant, due to large, relative to that in typical AMG use for partial differential equations, vertex degrees. Preconditioning for graph Laplacians is a competitive research area, but we have limited our testing to MPI software available via PETSc interfaces. 

\section{Results}\label{s:r}%

\begin{table}[]
\centering
\caption{Typical Python timing in sec for 100\% correct partition of static graphs from the Challenge Datasets with Known Truth}
\label{t:ts}
\begin{tabular}{|l|ccccc|}
 \hline
$|V|/|E|$ 	& 5K/.1M & 20K/.4M & 50K/1M & .5M/10M & 2M/41M \\
 \hline
\# Clusters 	& 19 	& 32 	& 44 	& 98 &160 \\
 \hline 
LOBPCG 	& $<$1 & 2 & 7 & 270 & 1700 \\
Base	& 400 & 5100 & 30000& N/A & N/A\\
 \hline
\end{tabular}
\end{table}
We have tested our MATLAB, Python, and C codes on several platforms, ranging from MS Windows and Mac laptops to r4.16xlarge AWS EC2 instances. Typical timing in seconds for 100\% correct partition of static graphs in Table~\ref{t:ts} is for a single node with 20 cores of Intel Xeon CPU E5-2660 v3 2.60GHz and 128 GB DDR4 RAM. We have not attempted to run the base Python code for graphs with over 50K vertices. 


Checking scalability as the size of the graph increases in Table \ref{t:ts}, we see that the timing grows at least linearly with the product of the number of edges $m=|E|$ and the number of clusters, we denote by $k$. We remind the reader that we simultaneously compute $l$ approximate eigenvectors in LOBPCG and keep $l$ slightly larger than the expected number $k$ of the clusters. LOBPCG performs linear algebraic operations with multi-vectors ($n$-by-$l$ matrices), at the cost proportional to $n\cdot l \approx n\cdot k$. 
In single-precision MATLAB tests, not reported here, but where we get more reliable timing measurement for graphs with 5K vertices, for $10$x growth from the graph size 5K/.1M  Vertices/Edges to 50K/1M Vertices/Edges at the same time increases $2.5$x the number of clusters from $19$ to $44$, resulting in approximately $25$x increase in computation time from $.3$ sec to $7$ sec. 
In Table \ref{t:ts}, $10$x growth from the graph size 50K/1M  Vertices/Edges to .5/10M Vertices/Edges at the same time increases $2.2$x the number of clusters $k$ from $44$ to $98$, resulting in approximately $40$x increase in computation time from $7$ sec to $270$ sec, so it appears that the growth may be quadratic in $k.$

Indeed, LOBPCG also computes Gram matrices for multi-vectors, where the computation time is proportional to $n\cdot l^2$. When the number $k\leq l$ of the clusters grows, this quadratic term in the costs may start dominating. However, we observe in Table \ref{t:ts} that $4$x growth from the graph size .5M/10M  Vertices/Edges to 2M/40M Vertices/Edges at the same time increases $1.6$x the number of clusters $k$ from $98$ to $160$, but resulting in only $6.3$x increase in computation time from $270$ sec to $1700$ sec, which is linear in $n\cdot k$. This effect may be explained by faster convergence for larger $k$ and efficient soft locking \cite{knyazev2004hard} of already converged eigenvectors in LOBPCG. 

The main memory usage in LOBPCG is for $6$ multi-vectors---matrices of the size $n$-by-$l$, in addition to whatever memory is needed to store the adjacency matrix $A$ or its equivalent, that is required to perform the multiplication of a multi-vector by the graph Laplacian. Simultaneous computing of $l=162$ eigenvectors for the graph with 2M/40M Vertices/Edges requires approximately 50 (25) GB of RAM in double (single) precision in our Python (MATLAB) codes. 

To save memory, LOBPCG can also perform hard-locking, i.e. compute eigenvectors one-by-one or block-by-block, in the constrained subspace that is complementary to all previously computed eigenvectors. Hard-locking is implemented in all LOBPCG versions, and is easy to use, but may lead to slower LOBPCG convergence. Even more radical memory saving approach is the standard recursive spectral bisection, which however would likely decrease both the convergence speed and the quality of partition. The Challenge graphs are small enough not to have forced the issue, allowing multi-way partitioning simply using the full block-size in LOBPCG without hard-locking, except for the largest graph with 5M vertices, when our double-precision code runs out of 128 GB of memory. 

Our main Python and MATLAB codes are very short, but call the LOBPCG function and a few standard functions for matrix computations, e.g., available in LAPACK. 

Accuracy of partitions for the streaming Partition Challenge Datasets with Known Truth Partitions is displayed using optimal partition matching (PM), pairwise recall (PR), and pairwise precision (PP). We report separately the results of streaming with random (R) and warm-start (W) initializations.

Our spectral clustering is generic, i.e., based entirely on the graph adjacency matrix, not making any assumptions specific to the block model, used to generate the graphs for the Challenge, in contrast to the baseline code. Unsurprisingly, PM and PP values for our partitions may at initial stages of graph streaming be small, relative to those for the base code, for small-size graphs, since the spectral gap may determine the number of clusters different from $k$.  We~report the correctness only at stages $5-10$ of the streaming process, except for 5K and 2M cases, where we provide the complete data.  

 The timing is reported in seconds in all tables. 
Using the warm-start (W) in LOBPCG compared to random (R) initializations not only commonly speeds up LOBPCG convergence, but has also typically positive effect on correctness of the partition, especially in the case of emerging edges.

Partitioning streaming graphs with snowballing effects brings new challenges, compared to the case of emerging edges. 
Since new vertices emerge in the streaming, we have to come up with the warm-start procedure, which gives us the values to initialize eigenvector components corresponding to the previously missing vertices for our spectral clustering that  
utilizes the vertex-based graph Laplacian. 

We resort to the trivial scenario, where the missing values are just chosen zero, while the previously computed values are reused wherever available. Such a warm-start (W) is beneficial in most tests, compared to random (R) initialization of all components of approximate eigenvectors. A simple alternative (not tested) is to randomize the missing enters.
We conjecture that the best results might be obtained via graph-based signal reconstruction techniques, such as in \cite{2017arXiv170503493K}, that interpolate a given signal on some vertices of a graph to all the vertices, using low-pass graph based filters.   

We finally note that numerical results somewhat vary when tests repeated even on the same hardware, due to random initialization. The presented results are typical, in our experience.

\begin{table}[]
\centering
\caption{5K Emerging (top) and Snowball (bottom) Base vs. LOBPCG}
\label{t:e5K}
\begin{tabular}{|c|c|ccc|c|ccc|}
 \hline
  \multicolumn{1}{|c|}{$|V|$}& \multicolumn{4}{c|}{Base}& \multicolumn{4}{c|}{LOBPCG warm-start}\\
  \hline
  \multicolumn{1}{|c|}{} & sec & PM& PR& PP& sec &  PM& PR& PP\\
  \hline
1  & 507 & .088 & 1 & .056      & .5 & .14	&.21	&.08\\
2  & 617 & .941 & .97 & .895   & .5 & .09	&.99	&.06\\
3  & 460 & .996 & .992 & .993  & .5 & .09	&.99	&.06\\
4  & 447 & .919 & .868 & .999 & .7  & .97	&.98	&.96\\
5  & 439 & 1 & 1 & 1              & .7 & .99	&.98	&1\\
6  & 446 & .919 & .869 & 1     & 1.1  & .97	&.99	&.96\\
7  & 442 & .926 & .871 & 1     & .7 & 1 & 1 & 1             \\
8  & 432 & 1 & 1 & 1              & .8 & 1 & 1 & 1             \\
9  & 415 & 1 & 1 & 1              & 1.1 & 1 & 1 & 1             \\
10 & 409 & .921 & .869 & 1     & 1.4 & 1 & 1 & 1            \\
\hline
1 &	22 	 &.550 & .382 & .656 	&	.2&	.34 & .999 & .14\\
2 &	52	 &.915 & .751 & .981 	&	.2&	.25 & .973 & .08\\
3 &	80	 &.999 & .999 & .999 	&	.2&	.93 & .929 & .99\\
4 &	113	 &1 & 1 & 1		&	.4 &	.13 & .988 & .06\\
5 &	146	 &1 & 1 & 1		&	.3 &	.27 & .985 & .08\\
6 &	190	 &1 & 1 & 1		&	.4 &	.26 & .999 & .08\\
7 &	244	 &1 & 1 & 1		&	.4 &	1 & 1 & 1\\
8 &	298	 &1 & 1 & 1		&	1.3 &	1 & 1 & 1\\
9 &	366	 &1 & 1 & 1		&	1 &	1 & 1 & 1\\
10 &	416	 &1 & 1 & 1		&	1.4 &	1 & 1 & 1\\
\hline
\end{tabular}
\end{table}

\begin{table}[]
\centering
\caption{2M Emerging (top) and Snowball (bottom) LOBPCG using Random (R) vs. Warm-start (W) initializations}
\label{t:e2M}
\begin{tabular}{|c|cc|cc|cc|cc|}
  \hline
  \multicolumn{1}{|c|}{\#} & secR&secW& PMR& PMW& PRR& PRW& PPR&PPW\\
  \hline
1&850 &800 &.016& .016 & .008	& .344&.008 &.007\\
2&1200&1000&.014 & .013 & .354	& .662& .007&.007\\
3&2100&1600&.019 & .013 & .784	& .916& .007&.007\\
4&3300&4700&.013 & .013 & 1	& 1& .007&.007\\
5&3300&2600&.013 & .013 & 1	& 1& .007&.007\\
6&2800&2800&.013 & .013 & 1	& 1& .007&.007\\
7&3000&1500&.013 & .964 & 1	& .993& .007&.943\\
8&2600&1200&1 	 & .992 & 1	& .999& 1&.991\\
9&3000&800 &1 	 & 1 & 1&1 &1 &1\\
10&3800&1600&1 	 & 1 & 1&1 &1 &1\\
\hline
1&440  & 420&   .08&.08 & .998&.998 & .014&.014\\
2&1020 & 960&  .037& .037& 1& 1&.008 &.008\\
3&1110 & 1400& .024& .083& 1& .794&.007 &.012\\
4&1450 & 1570&  .02& .02& .999& .98& .007&.007\\
5&1660 & 2000& .015&.015 & .999& .999& .007&.007\\
6&2260 & 2000& .037& .838& .993& .941& .007&.669\\
7&1980 & 720 & .036& .953& .999& .995& .007&.912\\
8&1840 & 1140&    1& .998& 1&1 & 1&.998\\
9&2400 & 710 &    1& 1& 1& 1& 1&1\\
10&3800& 3200&    1& 1& 1&1 & 1&1\\
\hline
\end{tabular}
\end{table}

\section{Conclusion}\label{s:c}
We review algorithms and software for eigenvalue problems, focusing on LOBPCG for spectral clustering and describing our spectral multi-way graph partitioning implemented in Python, MATLAB and C.
Our tests for all static graphs in the Challenge demonstrate 100\% correct partition, even though our spectral clustering approach is general, not tailored for the stochastic block model used to generate the graphs. Our~Python code is at least 100--1000 times faster for all tested graphs with 5K and more vertices compared to the baseline Python dense serial implementation of the reference method. 
Potential next steps include testing SLEPc, Trilinos/Anasazi, MAGMA, and CUDA versions of LOBPCG in shared and distributed memory environments for practical graph partitioning.






\bibliographystyle{IEEEtran}
\bibliography{refs} 
\newpage
\begin{table}[]
\centering
\caption{Emerging edges stages 5--10 LOBPCG Python}
\label{t:es5}
\begin{tabular}{|c|ccccc|}
\hline \multicolumn{6}{|c|}{Stage 5}\\
 \hline
  $|V|$ & 5K & 20K& 50K& .5M &2M\\
    \hline
sec R&.8 &5      &14     &443&3300\\
sec W&.7 &3      &4     &373&2600\\
  \hline
PM R&.99       &.965  &.892  &.021   &.013\\
PM W&.99       &.98  &.903  &.021   &.013\\
  \hline
PR R&.979       &.988  &.952  &.999   &1\\
PR W&.98      &.99  &.971  &.998  &1\\
  \hline
PP R&1.0       &.952  &.822  &.011   &.007\\
PP W&1.0     &.974  &.848  &.011 &.007\\
\hline
\hline \multicolumn{6}{|c|}{Stage 6}\\
 \hline
  $|V|$ & 5K & 20K& 50K& .5M &2M\\
    \hline
sec R&.7 &4      &15     &520&280\\
sec W&1.1 &1.3      &4     &290&2800\\
  \hline
PM R&.97       &.999  &.991  &.039   &.013\\
PM W&.97       &.983  &.983  &.917   &.013\\
  \hline
PR R&.988       &.997  &.997  &.998   &1\\
PR W&.988      &.998  &.996  &.979  &1\\
  \hline
PP R&.957       &1.0  &.987  &.012   &.007\\
PP W&.957     &.984  &.968  &.85 &.007\\
\hline
\hline \multicolumn{6}{|c|}{Stage 7}\\
 \hline
  $|V|$ & 5K & 20K& 50K& .5M &2M\\
    \hline
sec R&.9 &4.6      &10     &380&3000\\
sec W&.7 &1.4      &7     &280&1500\\
  \hline
PM R&1       &1  &.992  &.039   &.013\\
PM W&1       &1  &.992  &.987   &.964\\
  \hline
PR R&1       &1  &.997  &1.0   &1\\
PR W&1      &1  &.997  &.999  &.993\\
  \hline
PP R&1       &1  &.987  &.012   &.007\\
PP W&1     &1  &.987  &.982 &.943\\
\hline
\hline \multicolumn{6}{|c|}{Stage 8}\\
 \hline
  $|V|$ & 5K & 20K& 50K& .5M &2M\\
    \hline
sec R&.9 &3      &10     &360&2600\\
sec W&.8 &4      &4     &160&1200\\
  \hline
PM R&1       &1  &.992  &.997   &1\\
PM W&1       &1  &.992  &1   &.992\\
  \hline
PR R&1       &1  &.997  &.999   &1\\
PR W&1      &1  &.997  &1  &.999\\
  \hline
PP R&1       &1  &.987  &.995   &1\\
PP W&1     &1  &.987  &1 &.991\\
\hline
\hline \multicolumn{6}{|c|}{Stage 9}\\
 \hline
  $|V|$ & 5K & 20K& 50K& .5M &2M\\
    \hline
sec R&1 &4.5      &10     &480&3000\\
sec W&1 &1.4      &6     &260&800\\
  \hline
PM R&1       &1  &1  &1   &1\\
PM W&1       &1  &1  &1   &1\\
 \hline
PR R&1       &1  &1  &1   &1\\
PR W&1      &1  &1  &1  &1\\
 \hline
PP R&1       &1  &1  &1   &1\\
PP W&1     &1  &1  &1 &1\\
\hline
\hline \multicolumn{6}{|c|}{Stage 10}\\
 \hline
  $|V|$ & 5K & 20K& 50K& .5M &2M\\
    \hline
sec R&1.7 &4.8      &16     &800&3800\\
sec W&1.4 &5.6      &20     &600&1600\\
  \hline
PM R&1       &1  &1  &1   &1\\
PM W&1       &1  &1  &1   &1\\
 \hline
PR R&1       &1  &1  &1   &1\\
PR W&1      &1  &1  &1  &1\\
 \hline
PP R&1       &1  &1  &1   &1\\
PP W&1     &1  &1  &1 &1\\
\hline
\end{tabular}
\end{table}

\newpage
\begin{table}[]
\centering
\caption{Snowball stages 5--10 LOBPCG Python}
\label{t:sb5}
\begin{tabular}{|c|ccccc|}
\hline \multicolumn{6}{|c|}{Stage 5}\\
 \hline
  $|V|$ & 5K & 20K& 50K& .5M &2M\\
    \hline
sec R&.4 &1.3      &5.5     &280& 1660\\ 
sec W&.3 &1.2      &3.7     &200& 2000\\
  \hline
PM R&.268       &.963  &.337  &.02   &.015\\
PM W&.268       &.997  &.965  &.02   &.015\\
  \hline
PR R&.985       &.989  &.975  &.999   &.999\\
PR W&.985      &.993  &.989  &.999  &.999\\
  \hline
PP R&.082       &.933  &.049  &.011   &.007\\
PP W&.083     &1  &.946  &.011 &.007\\
\hline
\hline \multicolumn{6}{|c|}{Stage 6}\\
 \hline
  $|V|$ & 5K & 20K& 50K& .5M &2M\\
    \hline
sec R&.5 &1.6      &5.7     &310& 2260\\
sec W&.4 &.8      &4     &110& 2000\\
  \hline
PM R&.262       &.998  &.991  &.038   &.37\\
PM W&.262       &.998  &.979  &.946   &.838\\
  \hline
PR R&.999       &.996  &.996  &.999   &.993\\
PR W&.999      &.997  &.991  &.99  &.941\\
  \hline
PP R&.082       &1  &.986  &.012   &.007\\
PP W&.082     &1  &.967  &.908 &.669\\
\hline
\hline \multicolumn{6}{|c|}{Stage 7}\\
 \hline
  $|V|$ & 5K & 20K& 50K& .5M &2M\\
    \hline
sec R&.7 &2      &7.7     &270& 1980\\
sec W&.4 &1.3      &3     &200& 720\\
  \hline
PM R&1       &.998  &1  &.038   &.036\\
PM W&1       &.999  &.992  &.992   &.953\\
  \hline
PR R&1       &.996  &.999  &.1   &.999\\
PR W&1      &.999  &.997  &.999  &.995\\
  \hline
PP R&1       &1  &1  &.012   &.007\\
PP W&1     &1  &.987  &.988 &.912\\
\hline
\hline \multicolumn{6}{|c|}{Stage 8}\\
 \hline
  $|V|$ & 5K & 20K& 50K& .5M &2M\\
    \hline
sec R&.9 &2.3      &8.9     &320& 1840\\
sec W&1.3 &2      &4.5     &170& 1140\\
  \hline
PM R&1       &1  &1  &.997   &1\\
PM W&1       &1  &1  &.997   &.998\\
  \hline
PR R&1       &1  &1  &.999   &1\\
PR W&1      &1  &1  &.999  &1\\
  \hline
PP R&1       &1  &1  &.996   &1\\
PP W&1     &1  &1  &.996 &.998\\
\hline
\hline \multicolumn{6}{|c|}{Stage 9}\\
 \hline
  $|V|$ & 5K & 20K& 50K& .5M &2M\\
    \hline
sec R&1.2 &2.9      &9.8     &340& 2400\\
sec W&1 &5.5      &6.2     &380& 710\\
  \hline
PM R&1       &1  &1  &.997   &1\\
PM W&1       &1  &1  &.997   &1\\
  \hline
PR R&1       &1  &1  &.999   &1\\
PR W&1      &1  &1  &.999  &1\\
  \hline
PP R&1       &1  &1  &.996   &1\\
PP W&1     &1  &1  &.996 &1\\
\hline
\hline \multicolumn{6}{|c|}{Stage 10}\\
 \hline
  $|V|$ & 5K & 20K& 50K& .5M &2M\\
    \hline
sec R&1.4 &7      &18     &670& 3800\\
sec W&1.4 &5      &19     &550& 3200\\
  \hline
PM R&1       &1  &1  &1   &1\\
PM W&1       &1  &1  &1   &1\\
 \hline
PR R&1       &1  &1  &1   &1\\
PR W&1      &1  &1  &1  &1\\
 \hline
PP R&1       &1  &1  &1   &1\\
PP W&1     &1  &1  &1 &1\\
\hline
\end{tabular}
\end{table}

\newpage
\end{document}